\documentclass[a4paper,11pt]{article}
\pdfoutput=1 

\usepackage{jcappub} 
\usepackage{lineno}
\usepackage[T1]{fontenc} 
\title{\boldmath First direction sensitive search for dark matter with a nuclear emulsion detector at a surface site}

\author[a,1]{A. Umemoto\note{Corresponding author.},}
\author[b,c]{T. Naka,}
\author[d]{T. Shiraishi,}
\author[e,f]{O. Sato,}
\author[g,h]{T. Asada,}
\author[g,h]{G. De Lellis,}
\author[e]{R. Kobayashi,}
\author[g,h]{A. Alexandrov,}
\author[g]{V. Tioukov,}
\author[i]{N. D'Ambrosio,}
\author[j]{G. Rosa,}

\affiliation[a]{International Center for Quantum-field Measurement Systems for Studies of the Universe and Particles (QUP),
High Energy Accelerator Research Organization (KEK),\\ 1-1 Oho, Tsukuba, Ibaraki 305-0801,Japan}
\affiliation[b]{Department of Physics, Toho University,\\ 2-2-1 Miyama, Funabashi, Chiba 274-8510, Japan}
\affiliation[c]{Kobayashi-Maskawa Institute, Nagoya University,\\ Furo-cho, Nagoya, Aichi 464-8601, Japan}
\affiliation[d]{Department of Physics, Kanagawa University,\\ 3-27-1 Rokkakubashi, Kanagawa-ku, Yokohama, Kanagawa 221-8686, Japan}
\affiliation[e]{Department of Physics, Nagoya University,\\ Furo-cho, Nagoya, Aichi 464-8602, Japan}
\affiliation[f]{Institute of Materials and Systems for Sustainability, Nagoya University,\\ Furo-cho, Nagoya, Aichi 464-8601, Japan}
\affiliation[g]{Sezione INFN di Napoli, Complesso universitario di Monte S. Angelo ed. 6,\\ Via Cinthia - 80126, Napoli, Italy}
\affiliation[h]{Università di Napoli ``Federico II'', Dipartimento di Fisica "E. Pancini",\\ Via Cinthia - 80126 Napoli, Italy}
\affiliation[i]{Laboratori Nazionali del Gran Sasso (INFN),\\ Via G. Acitelli 22, 67100 Assergi L’Aquila, Italy}
\affiliation[j]{Sezione INFN di Roma, Piazzale Aldo Moro, 2 - 00185 Rome RM, Italy}

\emailAdd{aumemoto@post.kek.jp}

\abstract{Fine-grained nuclear emulsion films have been developed as a tracking detector with nanometric spatial resolution to be used in direction-sensitive dark matter searches, thanks to novel readout technologies capable of exploiting this unprecedented resolution. Emulsion detectors are time insensitive. Therefore, a directional dark matter search with such detector requires the use of an equatorial telescope to absorb the Earth rotation effect. We have conducted for the first time a directional dark matter search in an unshielded location, at the sea level, by keeping an emulsion detector exposed for 39 days on an equatorial telescope mount. The observed angular distribution of the data collected during an exposure equivalent to 0.59 g days agrees with the background model and an exclusion plot was then derived in the dark matter mass and cross-section plane: cross-sections higher than $1.3 \times 10^{-28}$ cm$^{2}$ and $1.7 \times 10^{-31}$ cm$^2$ were excluded for a dark matter mass of $10$ GeV$/c^2$ and $100$ GeV$/c^2$, respectively. 
This is the first direction sensitive search for dark matter with a solid-state, particle tracking detector.}

\begin{document}
\maketitle
\flushbottom

\section{Introduction}
\label{sec:intro}
Cosmological observations support the existence of dark matter, but no direct detection of dark matter has ever been made.
Dark matter candidates are new particles beyond the Standard Model, and understanding the nature of dark matter is one of the most fundamental problems in modern astroparticle physics. 
Weakly Interacting Massive Particle (WIMP) is the most popular candidate for dark matter. Direct detection experiments aiming at the observation of WIMP-nucleus elastic scattering can determine the WIMP mass and its interaction cross-section with Standard Model particles. 
Since the WIMP speed on the Earth reflects the solar system motion, two features appear in the nuclear recoil signal: a seasonal modulation of the energy spectrum and an angular anisotropy in the recoil direction. Different detector technologies have been developed worldwide to observe the seasonal modulation during the last decades~\cite{DAMA}~\cite{Xenon1T}~\cite{EDELWEISS}~\cite{CRESST}~\cite{Darkside}~\cite{Pico60}~\cite{NEWSG}.
Experiments with directional sensitivity can provide a more robust evidence than annual modulation thanks to the very good background separation; the angular distribution of signal is centered around the direction of Cygnus constellation, while the background distribution is expected to be isotropic~\cite{Spergel}~\cite{Mayet}~\cite{Hare}. 
However, due to the need for an excellent spatial resolution, the experiments currently being conducted with directional sensitivity are limited. The expected kinetic energy of the nuclear recoil signal is below 100 keV, which corresponds to a track length of 10 -100 nm for solid-state detectors and 0.1 - 1 mm for gas detectors. Low-pressure gaseous time projection chambers are useful for such track detection, and have become the most representative detection method for the directional dark matter search~\cite{Gas}.
The NEWAGE experiment comprising the use of CF$_4$ gas and $\mu$-PIC exhibited the highest sensitivity in the directional measurements~\cite{NEWAGE}. Other experiments such as CYGNO~\cite{CYGNO} and CYGNUS ~\cite{CYGNUS} are developing their gas detectors for directional measurements. 
On the other hands, solid-state tracking detectors with nanoscale spatial resolution are scalable in mass and can achieve higher sensitivity, with rather low energy thresholds, thus being very promising. 

The NEWSdm Collaboration has developed the super-fine-grained emulsion, nano imaging tracker (NIT) to detect sub-micrometer trajectories of charged particles with the finest spatial resolution of several tens of nanometers~\cite{NIT} \cite{NEWSdm2017efa}.
Silver bromide nano-crystals with a small amount of iodine additives, AgBr(I), dispersed in a gelatin binder are the sensitive element to detect a charged particle track as the line of silver grains produced from AgBr(I) crystals. 
In the past, reports on several developments were published, such as the method for producing NIT~\cite{NIT}, the construction of an optical reading system (PTS2)~\cite{PTS2}, the success of detecting carbon 30 keV tracks using an elliptical fitting method~\cite{DFT} and the demonstration of super-resolution imaging methods utilizing surface localized plasmon resonance~\cite{SRPIM1}~\cite{SRPIM2}~\cite{Alexandrov}.
In this paper, we report a directional dark matter search experiment using NIT at a surface laboratory. By using an equatorial telescope, the arrival direction of dark matter can be fixed in the Galactic coordinate system. Although NIT has no time resolution, the equatorial telescope allows to perform a directional dark matter search with a NIT detector with the same statistical advantages as when using a detector with real time information. We have been conducting the measurement for 39 days and we have analyzed the corresponding track angular distribution for WIMP detection.

\section{Detector concept}
\label{sec:two}

\begin{figure}[h]
 \centering
 \includegraphics[width=0.4\textwidth]{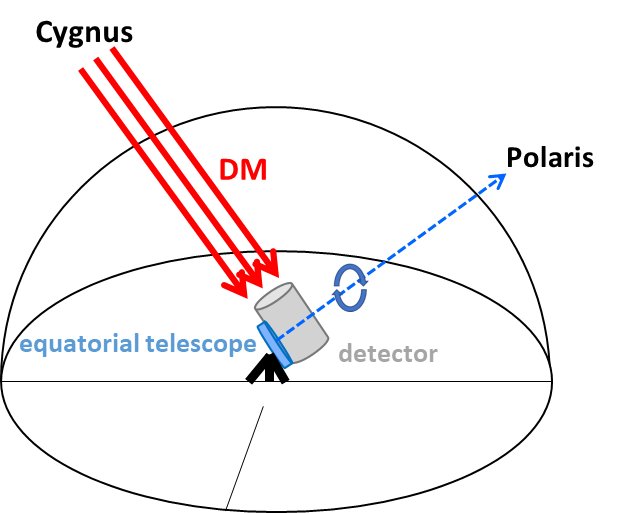}
\caption{Schematic of the apparatus using an equatorial telescope for directional dark matter search. }
\label{fig:Concept}
\end{figure}

The WIMP velocity on the Earth is affected by the solar system’s motion~\cite{Spergel}. 
The solar system is moving towards the Cygnus constellation, which then appears as the incoming direction of WIMPs when seen from the Earth.
In the laboratory system, the Cygnus direction changes with time due to the Earth rotation.
Detectors with time resolution can recompute the position of the Cygnus at the instant when the nuclear recoil event is produced, and the correlation between the Cygnus and the recoil angle can be evaluated event by event.
On the contrary, detectors without time resolution such as a NIT-based apparatus could detect only the integrated angular distribution of all events accumulated during the measurement period. 
Therefore, the number of events required to discriminate a flat background against a WIMP signal would increase by a factor between 1.5 and 3~\cite{TimeInteg}.
In order to compensate for this anisotropy loss, we have designed and constructed an apparatus with an integrated equatorial telescope. The equatorial telescope absorbs the Earth rotation effect, by orienting the detector towards the Cygnus constellation, with the orthogonal axis pointing to the Polaris and with the right ascension axis aligned with the Earth's axis, coping with the diurnal Cygnus motion. Figure~\ref{fig:Concept} shows the schematic view of the measurement system.

\section{Directional measurement}
\label{sec:three}
\subsection{Experiment setup}
\label{subsec:three1}
\begin{figure}[h]
 \centering
 \includegraphics[width=0.9\textwidth]{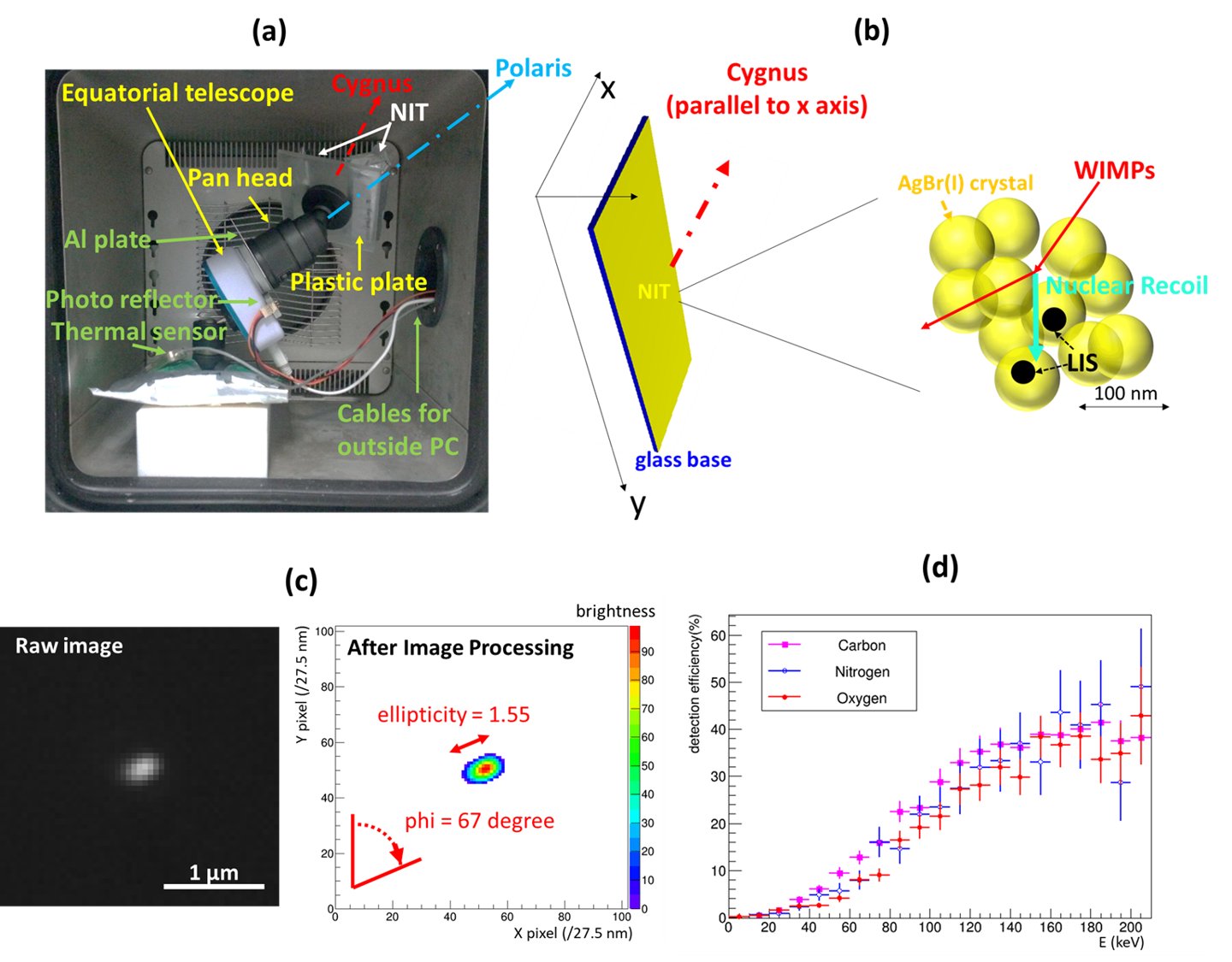}
\caption{(a) The experimental apparatus used in this measurement, consisting of: NIT films, an equatorial telescope mount, a pan head, an aluminum plate and a photo reflector. (b) NIT reference system with the Cygnus direction and NIT track detection mechanism. (c) Optical image of a carbon 100 keV track detected by the NIT and the PTS2 scanning system, and the computed parameters. (d) Detection efficiency of C, N, O recoils as a function of their kinetic energy. }
\label{fig:detectorsystem}
\end{figure}

The directional WIMP search was conducted for 39 days 
at the ground floor of the Science Building B in Nagoya University. The detector structure used in this measurement is shown in figure~\ref{fig:detectorsystem}(a).
Using a commercially available portable equatorial telescope mount ("NEW nano. tracker", SIGHTRON) and a pan head, one axis of the NIT defined as the X-axis was oriented along the direction of Cygnus as shown in figure~\ref{fig:detectorsystem}(b).
The base of the equatorial telescope mount was directed toward the Polaris, and the NIT installed at a tip of the pan head on the equatorial telescope mount was tilted in the Cygnus direction. A rectangular aluminum (Al) plate was installed between the pan head and the base of the equatorial telescope mount as a monitor by rotating with being synchronized the movement of the equatorial telescope. Since the long side of the rectangle Al plate was sometimes above a photo-reflector ("QTR-1A", SWITCH SCIENCE), we acquired the time when the reflected light began to be detected and monitored the rotation cycle of the equatorial telescope. The average cycle was 23 hours 56 minutes and 30 seconds during the measurement, in good agreement with the Earth’s rotation period.
This means that the NIT X-axis, set along the Cygnus direction, was kept within an angular deviation of 4 degrees during all the measurement. 

NIT sensitive layers used in this work had the following features: crystal size of 74.2 $\pm$ 8.9 (1 $\sigma$) nm, an average crystal density of 6.9 crystals/$\mu$m and a density of 3.1 $\pm$ 0.1 g/cm$^{3}$. 
It was poured on one side of a slide glass of 26 $\times$ 76 $\times$ 1 mm ("FRC-04", MATSUNAMI) to form a 40 $\mu$m-thick sensitive layer. 
Because at least two crystals are required for track detection, in principle tracks longer than 145 nm (the inverse of the crystal number density) can be detected. For carbon recoils, this is equivalent to the mean track length of a 45 keV monoenergetic recoil and to the maximum track length of a monoenergetic 20 keV recoil. 
The mass fraction of NIT consisted of Ag (41.5$\%$), Br (29.7$\%$), I (1.9$\%$), C (12.3$\%$), N (3.7$\%$), O (9.2$\%$) and H (1.8$\%$). These fractions have a 10$\%$ measurement uncertainty.
Given the typical features of nuclear recoil signals such as the track length and the amplitudes of cross-section for nuclear spin-independent interaction, C, N, and O recoils account for more than 95$\%$ of detectable nuclear recoils in the WIMP mass region between 10 to 100 GeV/c$^2$ ~\cite{Lewin}. We have then only considered those nuclei. 

When charged particles such as nuclear recoil excited AgBr(I) crystals by ionization during the measurement, silver cores referred to as “latent image specks” (LIS) were created in the crystals (figure~\ref{fig:detectorsystem}(b)). 
Through a standard chemical image development process using “metol–ascorbic acid” (MAA) after the measurement, the LIS grew to silver grains with complicated filament structures and sizes of 100 nm. Then we can observe charged particle tracks as a line of silver grains under an optical microscope. 
In order to maintain the sensitivity of AgBr(I) crystals and suppress fading effects, i.e. the disappearance of LIS due to the discharge, the entire detector was cooled down to -25 $\pm$ 1 $^{\circ}$C by using a constant temperature bath ("SH222", ESPEC) during the measurement. 
A reference sample, produced through the same process and developed soon after without any exposure at the beginning of the search, was used for comparison to estimate the background due to the production process.

An area of a few centimeters square (equivalent to $\sim$15 mg) near the center of NIT film was used for the analysis, discarding the thickness of 3 $\mu$m near the surface.
Optical images were obtained using an optical microscope denoted as PTS2~\cite{PTS2}, and elliptical parameters of optical image were calculated from the brightness distribution~\cite{DFT}.
Ellipticity and the direction of the major axis, denoted as phi, are used to estimate the track length and track angle, respectively. 
The spatial resolution of NIT determined from the crystal number density enables 3D axial tracking, but the PTS2 does not allow sufficient spatial resolution in the Z direction due to the depth of field. 
Therefore, 2D axial tracking was performed in this analysis. In the WIMP mass range mentioned above, there are less than 10$\%$ of the events with a recoil angle larger than 50$^{\circ}$ with respect to the z direction.
Therefore, accounting also for the detection mechanism, NIT get shrunk by a factor 0.6 with respect to the original thickness after image development, and a 2D analysis is appropriate to detect the track angle of nuclear recoil signals. 

Here, the elliptical parameter phi is defined as a two-dimensional polar vector in which the horizontal direction of the image is 90$^{\circ}$.
We set the sample to be same direction between phi = 90$^{\circ}$ and the X-axis in figure~\ref{fig:detectorsystem}(b). 
Figure~\ref{fig:detectorsystem}(c) shows an example of an optical image of a carbon 100 keV track detected by NIT using an ion implantation system.
This event has an ellipticity of 1.55 and phi = 67$^{\circ}$, respectively. The details of the measurement using the ion implantation system to evaluate the ellipse analysis are shown in~\cite{DFT}.

The ellipticity was required to be larger than 1.50 to avoid optical aberration and vibration effects in the PTS2 system. The detection efficiency of each C, N, O recoil with this ellipticity threshold is shown in figure~\ref{fig:detectorsystem}(d), estimated using a MC simulation based on SRIM and wave optics modeled from the measurement results using low-velocity carbon ion implantation. 
The efficiencies for carbon 30 keV and 100 keV were 3.8 $\pm$ 0.7$\%$ and 21.8 $\pm$ 2.8$\%$, respectively. Geometrical analysis of optical images selects events with a track length longer than 2.5 $\mu$m and they are removed in the signal candidates.

\subsection{Estimation of track angle distribution}
\label{subsec:three_2}
\subsubsection{WIMPs signal}
\label{subsubsec:three_2_1}
The angular distribution of the nuclear recoil signals expected in this measurement was calculated as follows.
Assuming the cross-section for spin-independent interactions of WIMP on nucleon, the recoil angle $\theta$ and recoil energy spectrum were obtained by referring to the methods in \cite{Lewin}\cite{Spergel}.
The cosmological parameters of dark matter are the following: the velocity distribution of dark matter in the milky way galaxy is assumed to be Maxwellian with an average value of 220 km/s, a relative earth velocity of 244 km/s and a galactic escape velocity of 650 km/s, respectively. The local dark matter density is assumed to be 0.3 GeV/cm$^{3}$.

\begin{figure}[h]
 \centering
 \includegraphics[width=1\textwidth]{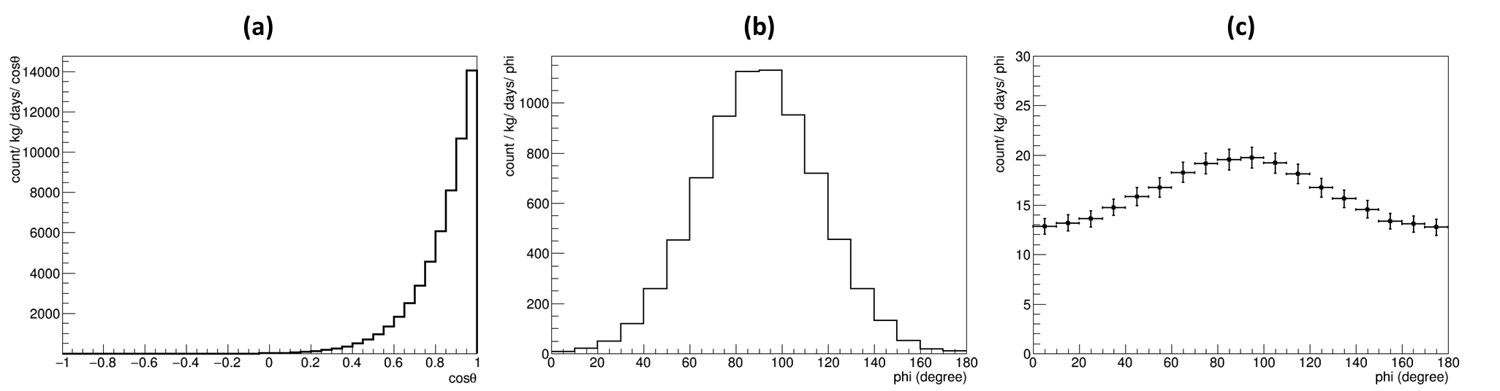}
\caption{Example of angular distribution of carbon recoils with energies above 30 keV induced by WIMP with mass of 10 GeV/c$^{2}$ and the spin-independent cross-section of 10$^{-32}$ cm$^{2}$.
(a) cos$\theta$ distribution. (b) 2D axial angle distribution derived from the cos$\theta$. The Cygnus direction corresponds to 90$^{\circ}$. (c) Expected phi distribution.}
\label{fig:angulardistribution}
\end{figure}

Defining the angle between the Cygnus direction and the carbon recoil as $\theta$, the cos$\theta$ distribution for recoil energies above 30 keV was obtained as shown in figure~\ref{fig:angulardistribution}(a). 
This was calculated with the assumption of WIMP with the mass of 10 GeV/c$^{2}$ and the spin-independent cross-section of 10$^{-32}$ cm$^{2}$. Since one axis of the 2D plane of the NIT is aligned with the Cygnus direction as shown in figure~\ref{fig:detectorsystem}(b), it is easy to covert the distribution from cos$\theta$ to 2D plane under the assumption that the velocity distribution of WIMPs on the earth is uniform in the remaining two axial directions. By a simple angular transformation, we obtained the 2D axial angle as shown in figure~\ref{fig:angulardistribution}(b), where the Cygnus direction corresponds to 90$^{\circ}$. In order to obtain the expected angular distribution, we estimated the effect of angular resolution and detection efficiency.

\begin{figure}[h]
 \centering
 \includegraphics[width=1\textwidth]{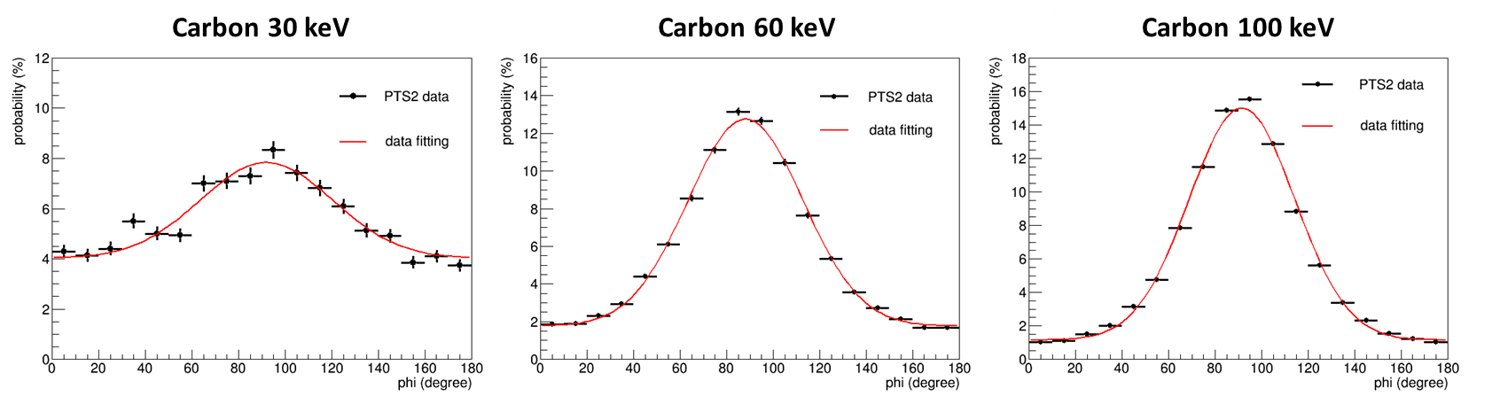}
\caption{Reconstructed angle phi distribution of the carbon 30, 60, and 100 keV ions with the ellipticity threshold of 1.5, respectively.}
\label{fig:carbon_angle}
\end{figure}

The angular resolution of the NIT was determined from the measurement of elliptical analysis for carbon ion tracks. Figure~\ref{fig:carbon_angle} shows the phi distribution of carbon 30, 60, and 100 keV tracks detected with ellipticity$\geq$1.5 selection. Each distribution has a peak at phi = 90 $^{\circ}$ which is the beam direction. 
The red solid lines in figure~\ref{fig:carbon_angle} are the fitting results using Eq.~(\ref{eq:Gaus_const}). The fitting parameters at each energy are summarized in table~\ref{tb:table1}.

\begin{equation}
 \frac{A}{\sqrt{2\pi}} exp (- \frac{(x -\mu)^{2}}{2\sigma^{2}}) + C
\label{eq:Gaus_const}
\end{equation}

\begin{table}[h]
\centering
\begin{tabular}{c|c|c|c}
\hline
carbon energy & A & $\sigma$ & C \\  
\hline
 30 keV & 273 $\pm$ 26 & 28.6 $\pm$ 2.3 & 4.02 $\pm$ 0.15 \\
\hline
 60 keV & 683 $\pm$ 8 & 24.8 $\pm$ 0.3 & 1.76 $\pm$ 0.04 \\
\hline
 100 keV & 791 $\pm$ 5& 22.8 $\pm$ 0.1 & 1.15 $\pm$ 0.02 \\
\hline
\end{tabular}
\caption{The angular fitting parameters using eq.~\ref{eq:Gaus_const} for the phi distribution of each energy in figure~\ref{fig:carbon_angle}.}
\label{tb:table1}
\end{table}

Since the ions were implanted with a monochromatic energy ($\pm$ 1 keV) and good angular uniformity ($<$ 1 mrad), the fitting function was used as an estimate of the phi probability density including the contribution from the angular resolution of NIT. The phi distribution expected in the directional WIMP search is reported in figure~\ref{fig:angulardistribution}(c). The uncertainty associated to each bin in figure~\ref{fig:angulardistribution}(c) was obtained by repeating 100 times the calculation of the phi distribution while accounting for the uncertainties of the detection efficiency and angular resolution. Although the distribution of signal has an isotropic term owing to the poor angular resolution, the peak is stably located at phi = 90 degree.

Since our measurements were carried out with carbon ions of 30, 60, and 100 keV, the phi distribution for other energies was evaluated by using the closest and lowest available energy. 
An energy threshold of 30 keV was used in the analysis, since the lowest energy used in the measurement was such. 
Moreover, since there is no significant difference between C, N and O recoils in the energy loss and angle dispersion due to multiple Coulomb scattering, the result obtained with the carbon ions were used for all nuclei. 
The phi distributions of C, N, and O recoils as a function of the WIMP mass were estimated.

\clearpage

\subsubsection{electron background}
\label{subsubsec:three_2_2}
\begin{figure}[h]
 \centering
 \includegraphics[width=0.5\textwidth]{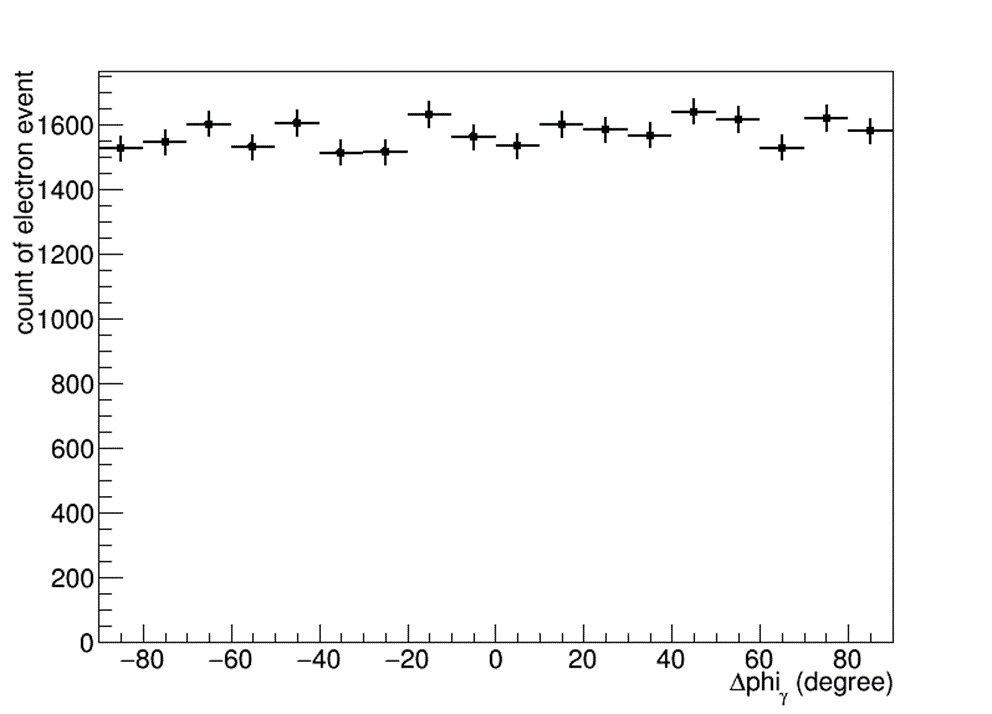}
\caption{$\Delta$phi$_{\gamma}$ distribution evaluated using an $^{241}$Am point source.}
\label{fig:gamma_angle}
\end{figure}

Since the detector system shown in figure~\ref{fig:detectorsystem}(a) was not shielded against environmental radiation, electron events due to the low energy environmental $\gamma$-rays are the leading background source.
$\gamma$-rays with energies below 100 keV produce electrons mainly via the Photoelectric effect.
The probability to reconstruct an electron event at a given phi angle was evaluated by using an $^{241}$Am $\gamma$-ray point source. 
The arrival direction of $\gamma$-rays was determined from the relative positions between the detected event and the center of radiation source, and the correlation between the arrival direction and phi was measured by the angular difference $\Delta$phi$_{\gamma}$.
The distribution of $\Delta$phi$_{\gamma}$ is shown in figure~\ref{fig:gamma_angle}.
We used the selection with ellipticity$\geq$1.5, which detected $\gamma$-rays from $^{241}$Am source with an efficiency of 1.2 $\pm$ 0.2 $\%$.
The distribution is flat with a p-value of 0.44.
Even if the arrival direction of environmental $\gamma$-rays has an angular dependence, the anisotropy is no longer seen in the detected phi distribution due to the scattering.

\subsection{Run result}
\label{subsec:three_3}
\begin{figure}[h]
 \centering
 \includegraphics[width=0.9\textwidth]{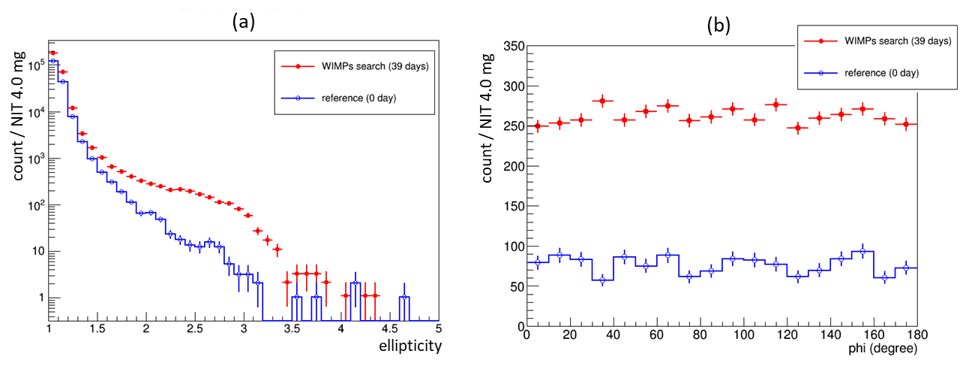}
\caption{Comparison of elliptical parameters of WIMP search for 39 days with the reference sample. (a) ellipticity distribution, (b) phi distribution with ellipticity$\geq$1.5.}
\label{fig:run_result}
\end{figure}

The data of the WIMP search for 39 days is shown in figure~\ref{fig:run_result}. 
For comparison, the data of a reference sample is also shown in figure~\ref{fig:run_result}.
Figure~\ref{fig:run_result} (a) and (b) show the ellipticity and phi distribution, respectively, normalized to the NIT mass of 4.0 mg. 
The analyzed mass of each sample is 15.1 mg for WIMP search and 3.9 mg for the reference sample, respectively.
From figure~\ref{fig:run_result}(a), a significant increase in the number of events with ellipticity$\geq$1.5 was detected in the WIMP search compared to the reference sample, and the detected event density was 372 $\pm$ 19 /mg for the reference sample and  1187 $\pm$ 25 /mg for WIMP search. 
An increase in the event density during the measurement was obtained as 21.2 $\pm$ 0.8 event/day/mg. This increase is consistent with the expected electron background.
The phi distribution of WIMP search shown in figure~\ref{fig:run_result}(b) does not show any peak in the Cygnus direction, corresponding to 90$^{\circ}$ in this plot. 
For the reference sample, this distribution shows only the accumulated background level during the production of the films, while the absolute phi value cannot be interpreted in terms of the direction to the Cygnus constellation. The phi distribution of WIMP search is compatible with a flat distribution with p-value of 0.14. 
No significant signals due to WIMPs was detected in this measurement.

\subsection{Angular distribution analysis}
\begin{figure}[h]
 \centering
 \includegraphics[width=0.9\textwidth]{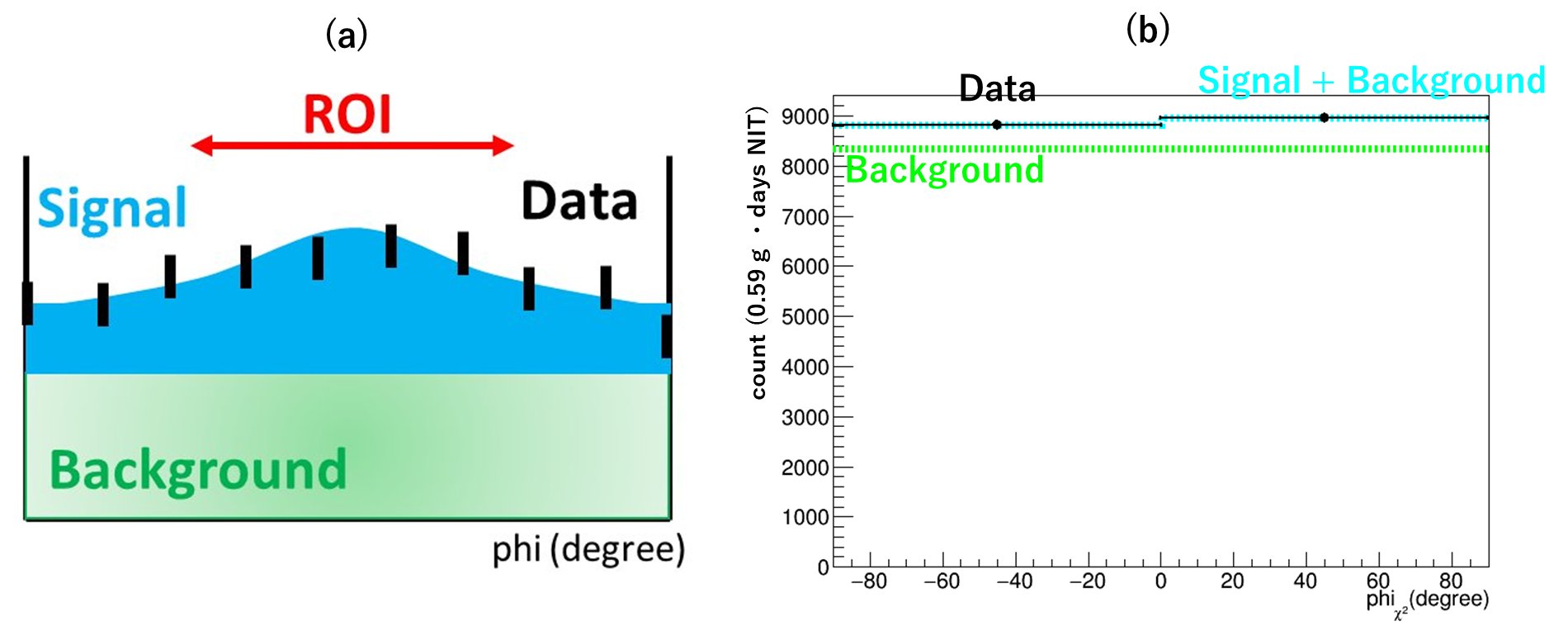}
\caption{(a) Expected phi distribution in a directional WIMP search. 
(b) The phi$_{\chi2}$ distributions for the data of the WIMP search and one case of the estimated S and B parameters, in the 10 GeV/c$^{2}$ WIMP mass assumption. }
\label{fig:angle_interupt}
\end{figure}

By analyzing the phi distribution of WIMP search, we constrained a parameter space of WIMP mass and spin-independent WIMP-nucleon cross-section.
Based on the conceptual diagram shown in figure~\ref{fig:angle_interupt}(a), we interpreted that the phi distribution consisted of WIMP signals (S) with a peak in the Cygnus direction and a background event (B) with a flat distribution. The angular distribution of signal was calculated as reported in section ~\ref{subsubsec:three_2_1}.
Given that the apparatus was not shielded, we expect that most of the background events come from environmental gamma rays. It has been confirmed in Sec~\ref{subsubsec:three_2_2} that low-energy gamma rays make a flat phi distribution. 
We treated the number of S and B as free parameters with the specified angular distributions mentioned above, and estimate the allowance of combinations of S and B which matches the experimental data by Chi-squared test.

In order to include the signal peak into the region of interest (ROI) and to realize a simple statistical analysis, the plot was made with only two bins, with a width of 90 degrees each. The right bin center is adjusted to the Cygnus direction, and this angular space is explained by using phi$_{\chi^{2}}$ which is a parameter relatively shifted phi. The phi$_{\chi^{2}}$ = 45 degree corresponds to phi = 90 degree. The black dot plotted in figure~\ref{fig:angle_interupt}(b) is phi$_{\chi^{2}}$ distribution of 17778 events detected in WIMP search with an exposure of 0.59 g$\cdot$ days (15.1 mg scanning mass times 39 days measurement).
For example, assuming the average detector performances, the combination of S =1090 and B =16688 is a best fit for WIMP mass of 10 GeV/c$^{2}$, as shown in figure~\ref{fig:angle_interupt}(b). 

Here, we have considered the experimental uncertainties reported in table~\ref{tb:uncertainty}.
The 20$\%$ error in the detection efficiency is due to the uncertainty of the MC simulation. The errors of the NIT density and mass fraction were derived from the measurement uncertainties. 
These uncertainties affect the amplitude of the signal distribution, which were categorized as "amplitude error". On the other hand, the angular resolution reported in table~\ref{tb:table1} and the accuracy of the Cygnus direction due to the equatorial telescope installation affect the shape of the signal distribution. They are categorized as "angle error".
Compared to these systematic errors, the statistical errors in the signal and background are negligible.

\begin{table}[h]
\centering
\begin{tabular}{c|c|c}
\hline
Source & Uncertainty & Categories \\ 
\hline
angular resolution & table1 & angle error\\
\hline
Cygnus direction & 10 degree & angle error\\
\hline
detection efficiency & 20$\%$ & amplitude error\\
\hline
density of NIT  & 3.2$\%$ & amplitude error\\
\hline
mass fraction of each C, N, O & 10$\%$ & amplitude error\\
\hline

\end{tabular}
\caption{Uncertainties considered in the analysis of this directional WIMP search.}
\label{tb:uncertainty}
\end{table}

Given the difficulty to produce an upper limit by accounting simultaneously for two uncertainties on a one-dimensional angular distribution, we have applied the following procedure. 
We call the bins reconstructed direction 1 and 2 (direction 2 includes the Cygnus direction) and the observed number of events in the experiment N$_{1}$ and N$_{2}$ (N = N$_{1}$ + N$_{2}$, N$_{1}$ = 8814 and N$_{2}$ = 8964). 
Signal and background components are S (S = S$_{1}$ + S$_{2}$,  S$_{1}$ $<$ S$_{2}$) and B (B = B$_{1}$ + B$_{2}$, B$_{1}$ =  B$_{2}$ = 0.5 $\times$ B), respectively.

Assuming the angle errors for signal distribution, we introduced a parameter S'.

\begin{align}
S' = \Delta_{1} \times S_{1} + \Delta_{2} \times S_{2} = S'_{1} + S'_{2} = S \\
\Delta_{2} = (S - \Delta_{1} \times S_{1})/ S_{2}         \end{align}

The $\Delta_{1}$ and $\Delta_{2}$ are smearing factors approximated by Gaussian with the center value of 1.0 for S$_{1}$ and S$_{2}$ when the angle errors were taken into account. 
The peak was still in Cygnus direction even if the uncertainties were considered. It was always satisfied that S'$_{1}$ $<$ S'$_{2}$ with the ratio of S'$_{2}$/S'$_{1}$ of 1.34 $\pm$ 0.07 (1$\sigma$) for WIMP mass of 10 GeV/c$^{2}$. 
To estimate the allowance of combinations of S and B, $\chi^{2}$ was calculated by using Eq.~(\ref{eq:Chi2_cal}). 

\begin{equation}
\chi^{2} = \frac{[N_{1}-(S'_{1} + B_{1})]^{2}}{S'_{1} + B_{1}} + \frac{[N_{2}-(S'_{2} + B_{2})]^{2}}{S'_{2} + B_{2}}
\label{eq:Chi2_cal}
\end{equation}
 
Then, probability density (Prob) value was calculated with a range up to $\chi^{2}$ = 18.5 in $\Delta \chi^{2}$ = 0.01 step to cover 99.99$\%$ in the case of degree of freedom (dof) of two.
If there were some combinations of S and B with the same $\chi^{2}$, Prob was divided by the number of combinations to keep the $\chi^{2}$ distribution. Thorough the process, a combination (S, B, Prob) can be derived.
Then, the amplitude errors were considered in order to obtain the true expectation value of detected signal (S$_{true}$) and background (B$_{true}$).
S$_{true}$ = $\Delta_{3}$ $\times$ S and B$_{true}$ = $\Delta_{4}$ $\times$ B, respectively.
The $\Delta_{3}$ and $\Delta_{4}$ account for the gaussian smearing of signal and background due to the amplitude uncertainty. Although we conducted the analysis without explicitly discussing the background source, for example, since electron recoils expected as a main background component were detected through the same fundamental process as nuclear recoils, we adopt a conservative approach by setting $\Delta_{3}$ $\sim$ $\Delta_{4}$.

\begin{figure}[h]
 \centering
 \includegraphics[width=1\textwidth]{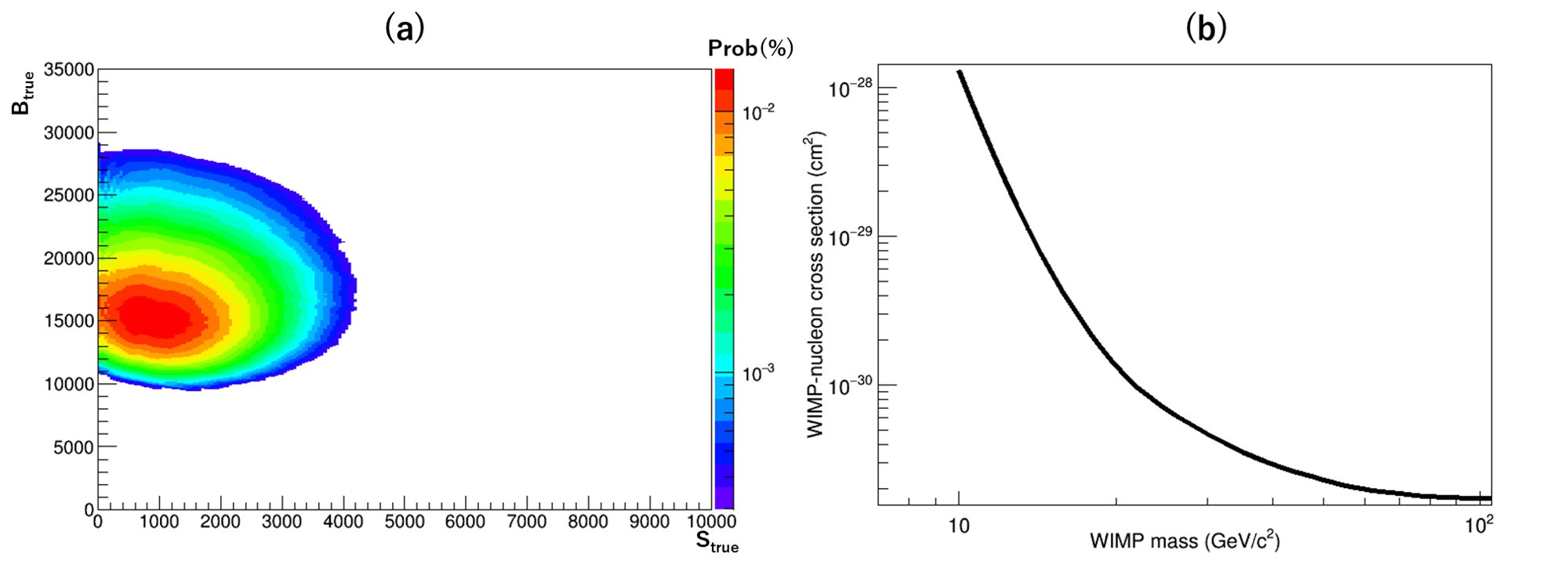}
\caption{(a) The probability distribution for the combination of S$_{true}$ and B$_{true}$ for a WIMP mass of 10 GeV/c$^{2}$ which matches experimental data with a confidence level of at least 95$\%$ C.L. or greater. 
(b) 95$\%$ C.L. upper limits on the spin-independent WIMP-nucleon cross-section as a function of the WIMP mass.}
\label{fig:run_sum}
\end{figure}

We conducted the calculations for 50000 times by randomizing $\Delta_{1}$ ($\Delta_{2}$) and $\Delta_{3}$ to average (S$_{true}$, B$_{true}$, Prob), and a list was made for S$_{true}$ and B$_{true}$ by ranking them according to the Prob.
Accumulating Prob starting from the largest one until the total reaches 95$\%$, the maximum value of S$_{true}$ was assumed to be the upper limit on the WIMP signal at 95$\%$ confidence level (C.L).
The probability distribution in space of  S$_{true}$ and B$_{true}$ is shown in figure~\ref{fig:run_sum}(a). Hence, S$_{true}$ = 4240 was obtained as the upper value, and the upper limit on the spin-independent WIMP-nucleon cross-section of  1.3 $\times $10$^{-28}$ cm$^{2}$ was obtained for WIMP mass of 10 GeV/c$^{2}$. This is the first directional measurement using a particle tracking detector for the WIMP mass below 20 GeV/c$^{2}$. 
The same procedure was repeated for different mass of WIMPs, and an exclusion limit curve on the cross-section as a function of the WIMP mass was calculated as shown in figure~\ref{fig:run_sum}(b).
The current best limit was 1.7 $\times $10$^{-31}$ cm$^{2}$ for the WIMP mass of 100 GeV/c$^{2}$. 
The sensitivity to WIMPs with masses above 100 GeV/c$^{2}$ will be estimated in the near future after assessing the detector performance for Ag, Br, and I recoils.

If directional information is not incorporated into the analysis, the increase in observed event between the WIMP search and reference samples would conservatively be assumed as a signal amount.  
Taking into account the amplitude errors as uncertainties with negligible statistical errors, the upper limit of S$_{true}$ at 95$\%$ C.L reaches 18000.
In conclusion, the directional analysis yielded sensitivity improvements of 4.3 and 6.5 times compared to the analysis without directional information for WIMP masses of 10 GeV/c$^{2}$ and 100 GeV/c$^{2}$, respectively.

\section{Discussion}
\label{sec:four}
\subsection{Effect of material attenuation}
\label{subsec:four1}
We have neglected the effect of the flux attenuation along the propagation path, although this effect can be relevant when the cross-section is as large as 10$^{-30}$cm$^{2}$~\cite{DMTheor1}. 
Similarly, given the attenuation induced by the bedrock, this cross-section region cannot be searched by underground experiments, thus increasing the importance of the results presented in this paper. 

\subsection{Estimation of background source}
\label{subsec:four2}
Another measurement using lead and copper shield for NIT was performed at the same place to study the background due to environmental $\gamma$-rays. 
Hereafter, this measurement is referred to as “BG study.” For constructing the shield, a tough-pitch copper plate with a thickness of 1 cm was placed inside each lead plate to shield not only environmental gamma rays but also K-shell X rays (E = 70 - 90 keV) from lead induced by cosmic ray muons. The lead plate was assembled with a thickness of 5.0 cm or more in the 4$\pi$ direction. 
Owing to this radiation shield, effective environmental $\gamma$-ray flux with energy of E= 20 - 200 keV was expected to be below 1$\%$.
In BG study, an increase in the event density with ellipticity$\geq$1.5 was obtained as 3.1 $\pm$ 0.3 event /day/mg, which was about 1/7 compared to the result of WIMP search 21.2 $\pm$ 0.8 event/day/mg. 

This decrease was not consistent to an estimation from the effective environmental $\gamma$-ray flux using the radiation shield.
Therefore, background events due to other radiation sources should be appeared in the BG study.
Intrinsic radioisotopes such as $^{14}$C contained in NIT and such as decay chains of $^{238}$U and $^{232}$Th, and $^{40}$K in the slide glass become the background sources.
Each radioactivity was determined from the measurements using accelerator mass spectrometry, germanium detector and inductively coupled plasma mass spectrometry. $^{14}$C of NIT was 21 $\pm$ 6 Bq/kg, and $^{238}$U, $^{232}$Th and $^{40}$K in slide glass are 3.0 $\pm$ 1.0 Bq/kg, 0.9 $\pm$ 0.1 Bq/kg, and 1.5 $\pm$ 0.2 Bq/kg, respectively. 
Most of the background events caused by these intrinsic radioisotopes are electrons from $\beta$-decay; neutron and $\gamma$-ray are negligible because they need to interact and create a nuclear or electron recoil background.
Since BG study was also conducted at the surface laboratory, ionizing electrons from cosmic ray muons could be background events. 
The cosmic ray muon flux was estimated to be 10$^{-2}$/s/cm$^{2}$ from the PARMA model~\cite{PARMA} assuming that cosmic ray muons with energy lower than 1 GeV can not able to reach NIT films due to the attenuation of the shielding materials and of the building. 
The contribution of environmental neutrons is negligible due to the small flux.
For these background sources including environmental $\gamma$-rays in BG study, electron production was calculated by Geant4 simulation and the background event ratio was estimated from the relation of energy loss and track range in NIT. 
This estimate is summarized in Table~\ref{tb:BG_estimate}. The total number of background event rate was estimated to be 3.50 $\pm$ 0.92 event/day/mg, consistent with that of BG study data. 

In order to reduce these electron background, we have begun developing a low radioactive NIT detector by changing the base material from slide glass to acrylic (e.g. cycloolefin polymer), and been promoting an underground experiment in LNGS with constructing an excellent shield for $\gamma$-rays. 

\begin{table}[h]
\centering
\begin{tabular}{c|c|c}
\hline
Source & Activity/Flux & background event rate (/mg/day) \\
\hline
$^{14}$C in NIT & 21 $\pm$ 6 Bq/kg & 0.11 $\pm$ 0.04\\
\hline
$^{40}$K in slide glass & 1.5 $\pm$ 0.2 Bq/kg & 0.09 $\pm$ 0.04 \\
\hline
$^{238}$U chain in slide glass & 3.0 $\pm$ 1.0 Bq/kg & 0.53 $\pm$ 0.21\\
\hline
$^{232}$Th chain in slide glass  & 0.9 $\pm$ 0.1 Bq/kg & 0.21 $\pm$ 0.09\\
\hline
environmental $\gamma$ ray & $\mathcal{O}$(0.01) /s/cm$^{2}$  & 0.21 - 0.55\\
\hline
cosmic-ray ($\mu^{\pm}) $  & $\mathcal{O}$(0.01) /s/cm$^{2}$ & 2.35 $\pm$ 0.70 \\
\hline
Total & & 3.50 $\pm$ 0.92\\
\hline

\end{tabular}
\caption{Estimation of background sources and event rates detected in BG run.}
\label{tb:BG_estimate}
\end{table}

\newpage
\section{Conclusion}
\label{sec:five}
We conducted the first directional sensitive dark matter search using a super fine-grained emulsion detector placed on an equatorial telescope mount at the sea level. 
One axis of the NIT plane was aligned with the Cygnus direction with an accuracy of 10 degree during the measurement, and the correlation between angular distribution of detected events and the arrival direction of dark matter was analyzed. The angular distribution was flat with a p-value of 0.14, and we could derive therefrom a 95$\%$ confidence level upper limit on the spin-independent WIMP-nucleon cross-section.
Cross sections higher than $1.7 \times 10^{-31}$ cm$^2$ were excluded for a dark matter mass of 100 GeV/c$^{2}$. This is the first search for dark matter with a solid-state, particle tracking detector.

\section*{Acknowledgements}
This work was supported by Japan Society for the Promotion of Science (JSPS) KAKENHI Grant-in-Aid for Scientific Research (A) 18H03699 and Grant-in-Aid for Scientific Research on Innovative Areas (Research in a proposed research area)19H05806.
This work was supported by "Advanced Research Infrastructure for Materials and Nanotechnology in Japan (ARIM)" of the Ministry of Education, Culture, Sports, Science and Technology (MEXT). Proposal Number F-19-NU-0006, F-21-NU-0065, JPMXP1223NU0227.






\begin{thebibliography}{99}

\bibitem{DAMA} R.~Bernabei et~al., \textit{The DAMA project: Achievements, implications and perspectives.}, \href{https://doi.org/10.1016/j.ppnp.2020.103810}{\textit{Prog. Part. Nucl. Phys.} \textbf{114} (2020) 103810} 

\bibitem{Xenon1T} E.~Aprile et~al., (XENON Collaboration) \textit{Dark Matter Search Results from a One Ton-Year Exposure of XENON1T.}, \href{https://link.aps.org/doi/10.1103/PhysRevLett.121.111302}{\textit{Phys. Rev. Lett.} \textbf{121} (2018) 111302} 

\bibitem{EDELWEISS} Q.~Arnaud et~al., (EDELWEISS Collaboration) \textit{Optimizing EDELWEISS detectors for low-mass WIMP searches.}, \href{https://link.aps.org/doi/10.1103/PhysRevD.97.022003}{\textit{Phys. Rev. D.} \textbf{97} (2018) 022003} 

\bibitem{CRESST} G.~Angloher et~al., \textit{Results on light dark matter particles with a low-threshold CRESST-II detector.}, \href{https://doi.org/10.1140/epjc/s10052-016-3877-3}{\textit{Eur. Phys. J. C.} \textbf{76} (2016) 25} 

\bibitem{Darkside} P.~Agneset et~al., (DarkSide Collaboration) \textit{DarkSide-50 532-day dark matter search with low-radioactivity argon.}, \href{https://doi.org/10.1103/PhysRevD.98.102006}{\textit{Phys. Rev. D.} \textbf{98} (2018) 102006} 

\bibitem{Pico60} C.~Amole et~al., (PICO Collaboration) \textit{Dark matter search results from the complete exposure of the PICO-60 C$_{3}$F$_{8}$ bubble chamber.}, \href{https://link.aps.org/doi/10.1103/PhysRevD.100.022001}{\textit{Phys. Rev. D.} \textbf{100} (2019) 022001} 

\bibitem{NEWSG} Q.~Arnaud et~al., \textit{First results from the NEWS-G direct dark matter search experiment at the LSM.}, \href{https://doi.org/10.1016/j.astropartphys.2017.10.009}{\textit{Astropart. Phys.} \textbf{97} (2018) 54} 

\bibitem{Spergel} D.~ N.~Spergel, \textit{Motion of the Earth and the detection of weakly interacting massive particles.}, \href{https://link.aps.org/doi/10.1103/PhysRevD.37.1353}{\textit{Phys. Rev. D.} \textbf{37} (1988) 6}.

\bibitem{Mayet} F.~Mayet et~al., \textit{A review of the discovery reach of directional Dark Matter detection.}, \href{https://doi.org/10.1016/j.physrep.2016.02.007}{\textit{Phys. Rep.} \textbf{627} (2016) 20}.

\bibitem{Hare} C.A.J.~O'Hare et~al., \textit{Readout strategies for directional dark matter detection beyond the neutrino background.}, \href{https://link.aps.org/doi/10.1103/PhysRevD.92.063518}{\textit{Phys. Rev. D.} \textbf{92} (2015) 063518}.

\bibitem{Gas} G.~Sciolla and C.~J.~Martoff, \textit{Gaseous dark matter detectors.}, \href{https://dx.doi.org/10.1088/1367-2630/11/10/105018}{\textit{New J. Phys.} \textbf{11} (2009) 105018}.

\bibitem{NEWAGE} T.~Ikeda et~al., \textit{Direction-sensitive dark matter search with the low-background gaseous detector NEWAGE-0.3b".}, \href{ https://doi.org/10.1093/ptep/ptab053}{\textit{Prog. Theor. Exp. Phys.} \textbf{2021} (2021) 063F01}.

\bibitem{CYGNO} F.~D.~Amaro et~al., \textit{The CYGNO experiment, a directional detector for direct Dark Matter searches.}, \href{ https://doi.org/10.1016/j.nima.2023.168325}{\textit{Nucl. Instrum. Methods A.} \textbf{1054} (2023) 168325}.

\bibitem{CYGNUS} K.~Miuchi, E.~Baracchini, G.~Lane, N.~J.~C.~Spooner, S.~V.~Vahsen, \textit{CYGNUS.}, \href{https://dx.doi.org/10.1088/1742-6596/1468/1/012044}{\textit{J. Phys.: Conf. Ser.} \textbf{1468} (2022) 012044}.

\bibitem{NIT} T.~Asada, T.~Naka, K.~Kuwabara, M.~Yoshimoto, \textit{The development of a super-fine-grained nuclear emulsion.}, \href{https://doi.org/10.1093/ptep/ptx076}{\textit{Prog. Theor. Exp. Phys.} \textbf{2017} (2017) 063H01} 

\bibitem{NEWSdm2017efa} N.~Agafonova et~al., \textit{Discovery potential for directional Dark Matter detection with nuclear emulsions.}, \href{https://doi.org/10.1140/epjc/s10052-018-6060-1}{\textit{Eur. Phys. J. C.} \textbf{78} (2018) 758}


\bibitem{PTS2} T.~Katsuragawa, A.~Umemoto, M.~Yoshimoto, T.~Naka, T.~Asada, \textit{New readout system for submicron tracks with nuclear emulsion.}, \href{https://dx.doi.org/10.1088/1748-0221/12/04/T04002}{\textit{J. Instrum.} \textbf{12} (2017) T04002} 

\bibitem{DFT} A.~Umemoto et~al., \textit{Optical shape analysis based on discrete Fourier transform and second-order moment analysis of the brightness distribution for the detection of sub-micron range tracks in nuclear emulsion.}, \href{ https://doi.org/10.1093/ptep/ptaa132}{\textit{Prog. Theor. Exp. Phys.} \textbf{2020} (2020) 103H02}

\bibitem{SRPIM1} A.~Umemoto, T.~Naka, A.~Alexandrov, M.~Yoshimoto, \textit{Super-resolution plasmonic imaging microscopy for a submicron tracking emulsion detector}, \href{https://doi.org/10.1093/ptep/ptz033} {\textit{Prog. Theor. Exp. Phys.} \textbf{2019} (2019) 063H02}

\bibitem{SRPIM2} A.~Alexandrov, et~al., \textit{Super‐resolution high‐speed optical microscopy for fully automated readout of metallic nanoparticles and nanostructures}, \href{https://doi.org/10.1038/s41598-020-75883-z} {\textit{Sci. Rep.} \textbf{10} (2020) 18773}.

\bibitem{Alexandrov} A.~Alexandrov, T.~Asada, F.~Borbone, V.~Tioukov, G.~De~Lellis, \textit{Super-resolution imaging for the detection of low-energy ion tracks in fine-grained nuclear emulsions,}, [\href{https://arxiv.org/pdf/2304.03645.pdf}{\texttt{arXiv:2304.03645}}].


\bibitem{TimeInteg} C.A.J.~O'Hare, B. J.~Kavanagh, A. M.~Green, \textit{Time-integrated directional detection of dark matter.}, \href{https://doi.org/10.1103/PhysRevD.96.083011}{\textit{Phys. Rev. D.} \textbf{96} (2017) 083011} 

\bibitem{Lewin} J.~D.~Lewin and P.~F.~Smith, \textit{Review of mathematics, numerical factors, and corrections for dark matter experiments based on elastic nuclear recoil.}, \href{https://doi.org/10.1016/S0927-6505(96)00047-3}{\textit{Astropart. Phys.} \textbf{6} (1996) 87}.


\bibitem{DMTheor1} W.~L.~Xu, C.~Dvorkin, A.~Chael, \textit{Probing sub-GeV dark matter-baryon scattering with cosmological observables.} \href{https://doi.org/10.1103/PhysRevD.97.103530}{\textit{Phys. Rev. D.} \textbf{97} (2018) 103530} 

\bibitem{PARMA} T.~Sato, \textit{Analytical Model for Estimating the Zenith Angle Dependence of Terrestrial Cosmic Ray Fluxes}, \href{https://doi.org/10.1371/journal.pone.0160390}{\textit{PLoS ONE.} \textbf{11} (2016) 8}.




\end{thebibliography}
\end{document}